\documentclass{PoS}

\title{Molecular Clouds as Cosmic Ray Laboratories}

\ShortTitle{}

\author{{Sabrina Casanova}\\
        Institut für Theoretische Physik, Lehrstuhl IV: Weltraum und 
Astrophysik, Ruhr-Universität Bochum, 44780, Bochum, Germany \\
and \\
 Max Planck f\"ur 
Kernphysik, Saupfercheckweg 1, 69117, Heidelberg, Germany\\
        E-mail: \email{sabrina@tp4.rub.de, Sabrina.Casanova@mpi-hd.mpg.de}}

\author{{Felix A. Aharonian$^{1,2}$, Yasuo Fukui$^{3}$, Stefano Gabici$^{4}$, David I. Jones$^{1}$, Akiko Kawamura$^{3}$, Toshikazu Onishi$^{3}$, Gavin Rowell$^{5}$, Hidetoshi Sano$^{3}$, Kazufumi Torii$^{3}$, Francesca Volpe$^{1}$, Hiroaki Yamamoto$^{3}$}\\
$^{1}${Max Planck f\"ur Kernphysik, Saupfercheckweg 1, 69117, Heidelberg}\\
$^{2}${Dublin Institute for Advance Physics,31 Fitzwilliam Place, Dublin 2, Ireland}\\
$^{3}${Nagoya University,Furo-cho, Chikusa-ku, Nagoya City, Aichi Prefecture, Japan}\\
$^{4}${Laboratoire APC, 10, rue Alice Domon et L\'{e}onie Duquet, 75013 Paris, France}\\
$^{5}${School of Chemistry and Physics, University of Adelaide, Adelaide 5005, Australia}}
   
\abstract{ We will here discuss how the gamma-ray emission from molecular clouds can be used to probe the cosmic ray flux in distant regions of the 
Galaxy and to constrain the highly unknown cosmic ray diffusion coefficient. In particular we 
will discuss the GeV to TeV emission from runaway cosmic rays penetrating molecular clouds close to young and old supernova remnants 
and in molecular clouds illuminated by the background cosmic ray flux.
}

\FullConference{25th Texas Symposium on Relativistic Astrophysics - TEXAS 2010\\
		December 06-10, 2010\\
		Heidelberg, Germany}

\begin{document}

\section{The standard model of cosmic rays}

One hundred years after their discovery by the Austrian physicist Victor Hess, 
the origin of cosmic rays (CRs) is still unclear. Diffusive shock acceleration in supernova remnants (SNRs) is the 
most widely invoked paradigm to explain the Galactic cosmic ray spectrum. Galactic 
SNRs provide, in fact, the necessary power to sustain the Galactic cosmic ray population \cite{Ginzburg}. 
The direct observation of cosmic rays from the 
candidate injection sites such as supernova remnants is not possible since CRs escape the 
acceleration sites and eventually propagate into the Galactic magnetic fields. 
CR secondary data suggest that 
cosmic protons and nuclei diffuse in the magnetic fields for timescales of the order of about $t_{escape} \approx {10}^7 \, 
{(\frac{E}{10 \, {\rm GeV}})}^{-0.65} \, \rm{years}$, $E$ being the particle energy, 
before escaping the Galaxy. During these timescales the particles from individual sources lose memory of their origin, and contribute to the 
bulk of Galactic cosmic rays known as cosmic ray background or CR {\it sea}, losing the information 
on the original acceleration locations and spectra. If $h$ is the distance that CRs have to travel before 
escaping the Galaxy, then the diffusion coefficient will be $D \approx  h^2/t_{escape}$ and 
one deduces an energy dependence of the propagation in 
the Galactic disk, $D(E) = D_{10} \, {(\frac{E}{10 \rm{GeV}})}^{\delta}$, 
where the diffusion coefficient at 10 GeV is about $D_{10}={10}^{28} \, {\rm \frac{{cm}^2}{s}}$, and $\delta=0.3-0.7$. 
The average cosmic ray density is thus determined by the contribution of all Galactic sources over a 
long period of time comparable to the CR escape time from the Galaxy.

The information on the locations of individual CR sources, their spectra and their injection rate, which get lost 
during the diffusion and convection processes CRs undergo, can be traced 
back through the gamma-rays which cosmic rays radiate when they interact with the ambient gas in the 
interstellar medium. In fact, gamma-rays are emitted through decay of neutral 
pions produced in inelastic collision of 
CR hadrons and interstellar gas. Contrary to CRs, the gamma rays, being neutral, 
travel in straight lines from the site where they were 
emitted to the detector. For this reason gamma-ray astronomy 
has always played a key role to probe the Galactic cosmic ray flux and to solve the 
longstanding question of the origin of cosmic rays. 

If the bulk of Galactic CRs up to at least PeV energies are indeed 
accelerated in SNRs, then TeV gamma-rays are expected to be emitted during 
the acceleration process CRs undergo within SNRs \cite{Drury}. Indeed 
TeV gamma-rays have been detected from the shells of SNRs, such as RX~J1713.7-3946 \cite{Aharonian:nature}. 
However, such 
observations do not constitute a definitive proof that CRs are 
accelerated in SNRs, since the observed
emission could be produced by energetic electrons up scattering low energy photon fields. 
Gamma-rays are also expected to be emitted when the accelerated 
CRs propagate into the interstellar medium (ISM) \cite{Montmerle,Issa,Aharonian:1996,Gabici2009}. 
Whereas the atomic hydrogen is 
uniformly distributed in the Galaxy, the molecular hydrogen is usually aggregated in 
dense clouds, and the gamma-ray emission from molecular clouds is particularly intense and localised. 
We will here discuss how the gamma-ray emission from these molecular clouds can be used to probe the cosmic ray flux in distant regions of the 
Galaxy and to constrain the highly unknown cosmic ray diffusion coefficient. In Section \ref{sec:youngSNR} 
we will present the results of the modeling of the gamma-ray emission in molecular clouds close 
to young SNRs. In Section \ref{sec:w28} the broadband emission around the old SNR W28 is used to constrain the diffusion coefficient in the region.  In Section \ref{sec:background} we will describe a methodology to 
investigate the CR flux and spectrum in distant regions of the Galaxy from observations 
of gamma-rays emitted by CRs propagating through molecular clouds far from known CR sources. 
Our conclusions are given in Section \ref{sec:conclusions}.

\section{Looking for evidence of runaway CRs from SNRs}\label{sec:youngSNR}

Cosmic rays escaping supernova remnants diffuse in the
interstellar medium and collide with the ambient atomic and molecular gas \cite{Montmerle,Issa,Aharonian:1996,Gabici2009}.
From such collisions gamma-rays are created, which can possibly provide
the first evidence of a parent population of runaway cosmic rays. 
Before being isotropised by the Galactic magnetic fields, the injected CRs produce, in fact, $\gamma$-ray emission, 
which can significantly differ from the emission of the SNR itself, as 
well as from the diffuse emission contributed by the background CRs and electrons, 
because of the hardness of the runaway CR spectrum, which is not 
yet steepened by diffusion. These diffuse sources are often correlated with dense molecular clouds (MCs), which 
act as a target for the production of gamma-rays due to the enhanced local CR injection spectrum. SNRs are located in star forming regions, 
which are rich in molecular hydrogen. In other words, CR sources and MCs 
are often associated and target-accelerator systems are not 
unusual within the Milky Way \cite{Casse,Montmerle}. 
\begin{figure}
\centering
  \includegraphics[width=0.23\textwidth]{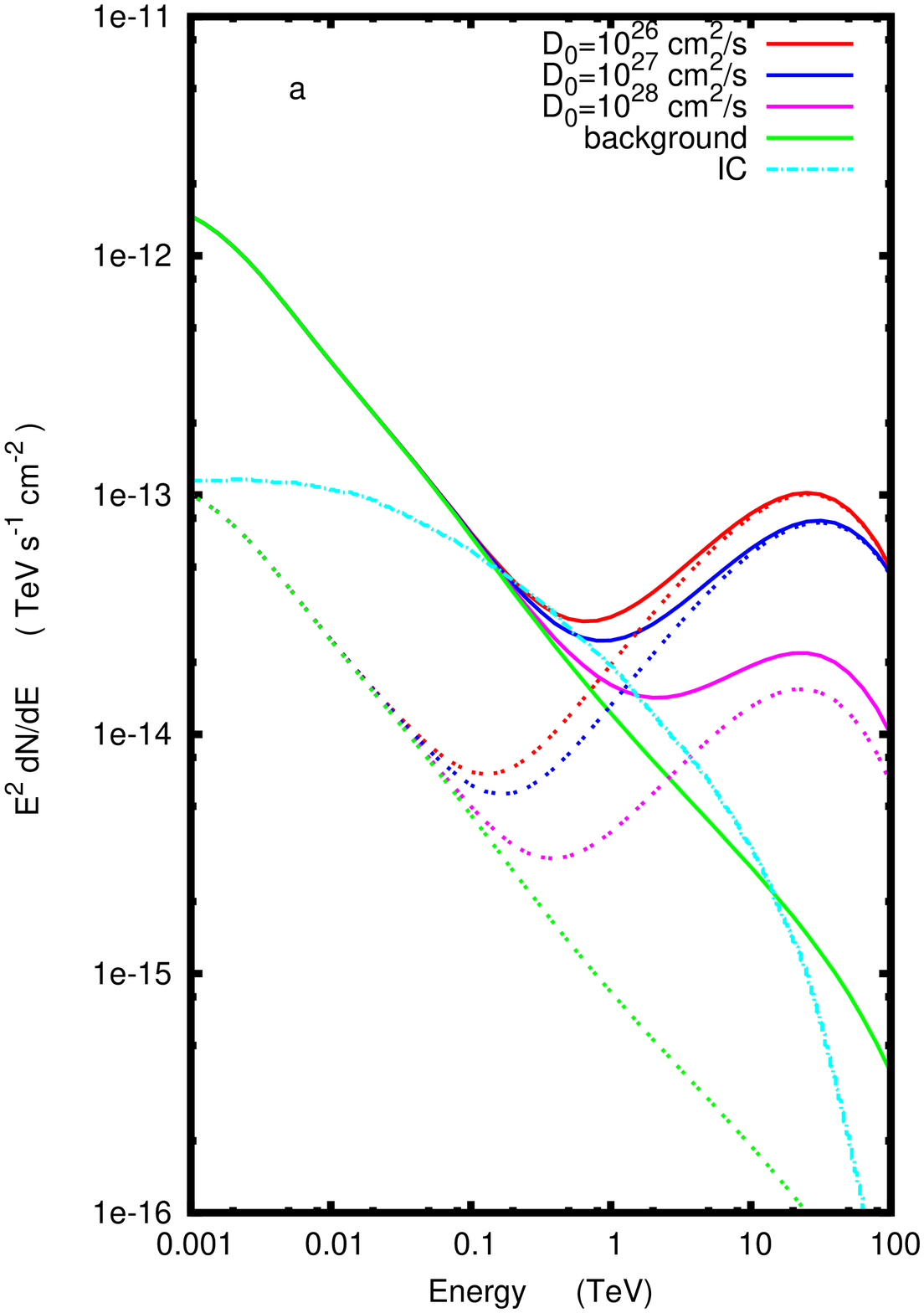}
  \includegraphics[width=0.23\textwidth]{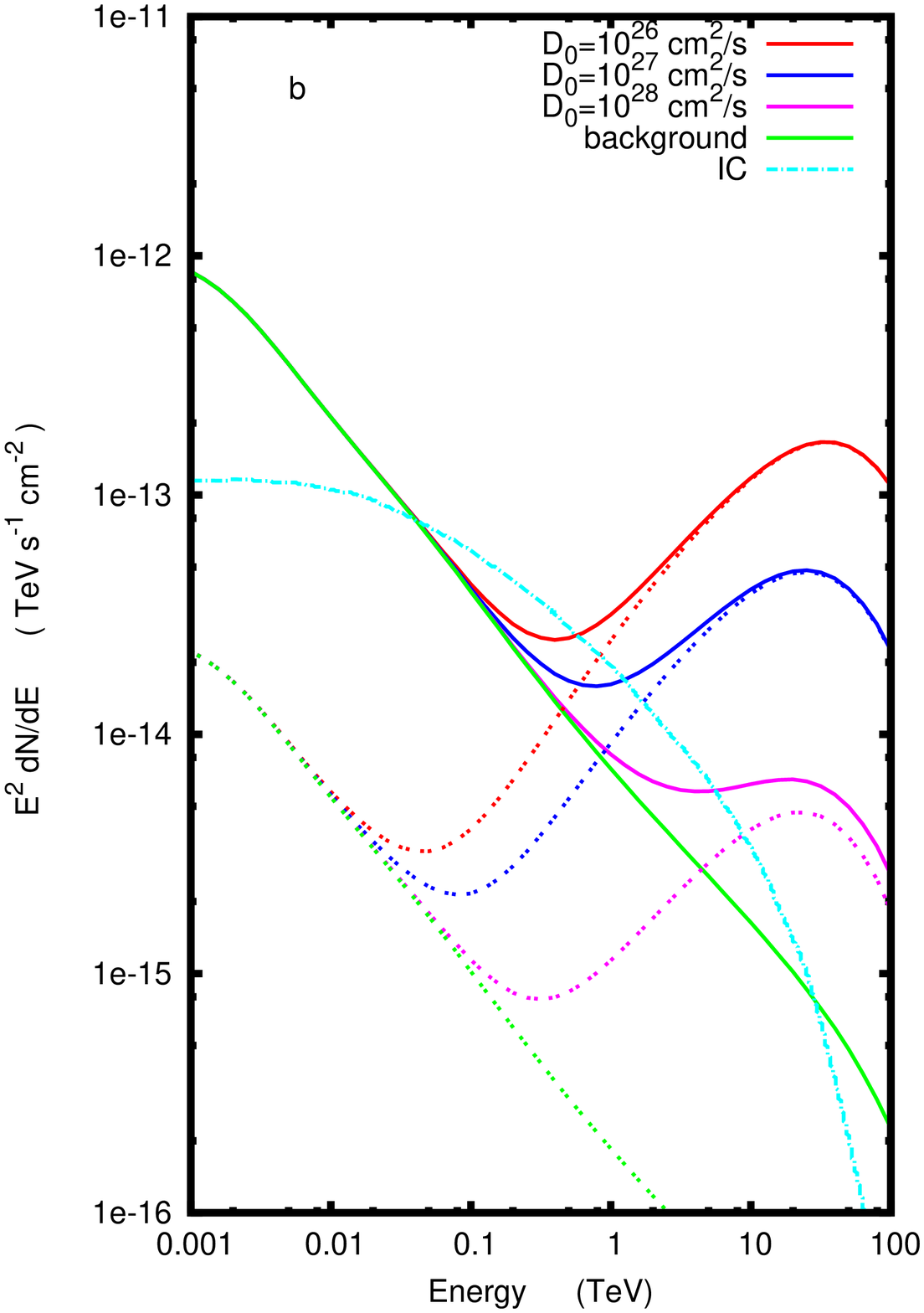}
  \includegraphics[width=0.23\textwidth]{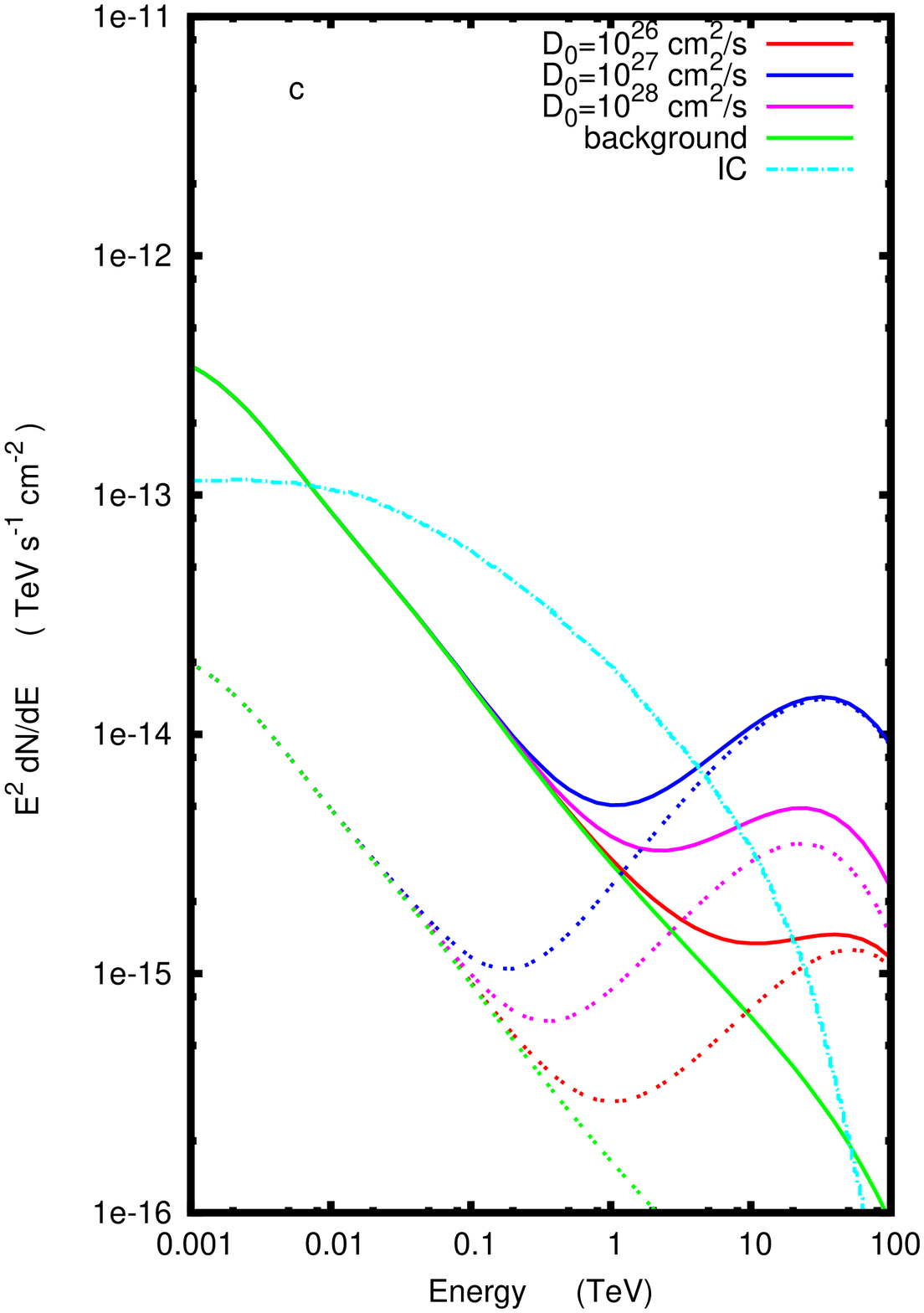}
  \includegraphics[width=0.23\textwidth]{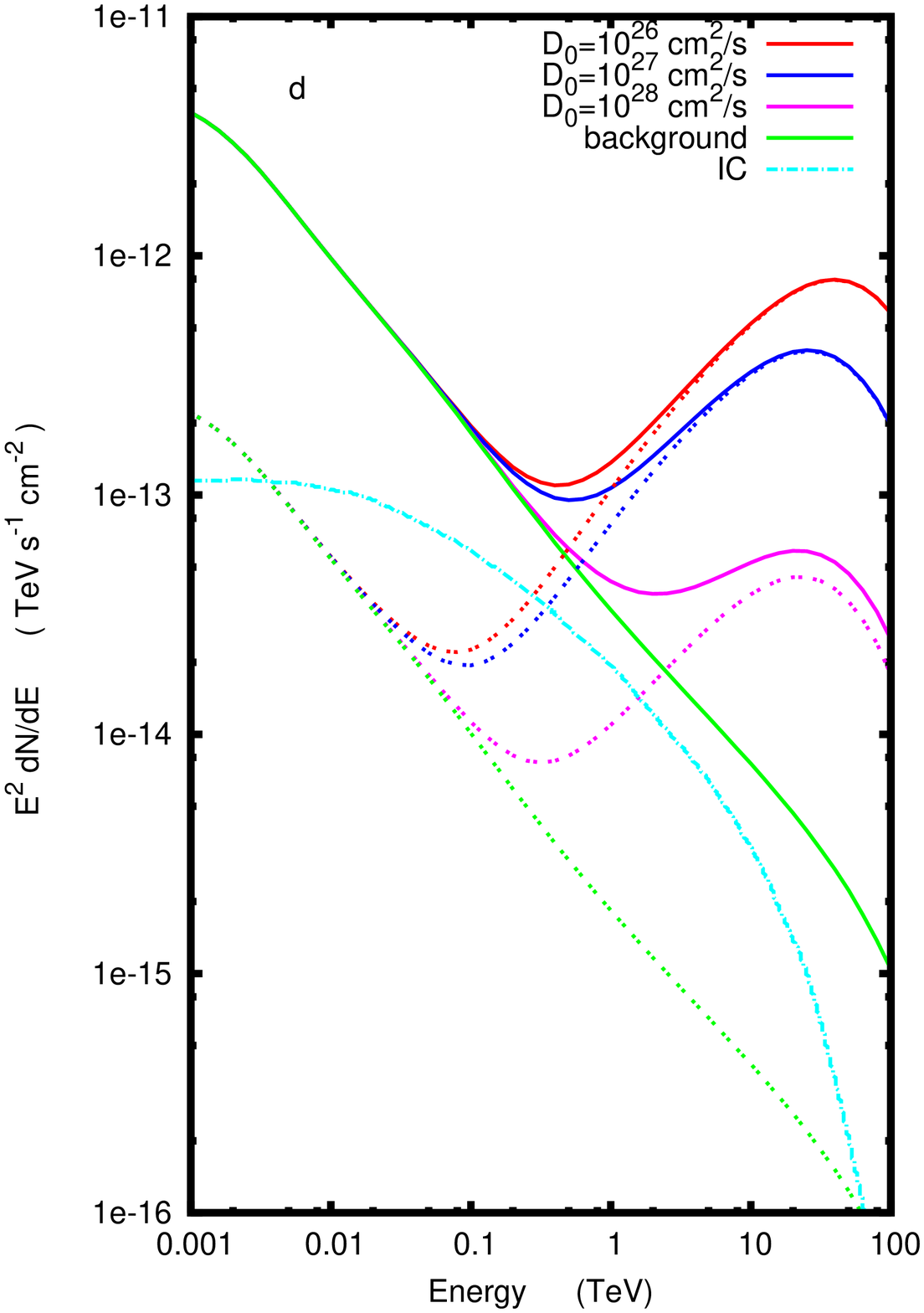}
\caption{The $\gamma$-ray energy flux in four different regions of 0.2 $\times$ 0.2 degrees 
around the positions a = (346.8, -0.4), b = (346.9, -1.4), c = (347.1, -3.0) and d = (346.2, 0.2). 
The emission produced along the line of sight distance between 
900 and 1100 parsecs is plotted in dashed lines, while the emission 
obtained by summing the radiation contributions over the whole line of sight distance, from 50 parsecs to 30000 parsecs, is 
shown in solid lines. The emission in the panels a,b,c and d, corresponding to the different locations, is 
plotted for different diffusion coefficients $D_0$. The emission from background 
CRs is also shown in each panel for comparison. The contribution to the emission from inverse Compton scattering of 
background electrons is 
indicated with a dashed light blue line. \label{fig66}}
\end{figure}
Figure \ref{fig66} shows the predicted gamma-ray spectra from hadronic interactions 
of background and runaway CRs in four regions around the SNR RX~J1713.7-394 \cite{Casanova2010b}. 
In all four locations the  hadronic gamma-ray emission is enhanced with respect 
to the hadronic emission due only to background CRs at energies above few TeVs 
as a consequence of the fact that the CR fluxes are enhanced above 100 TeV. CRs 
of about 100 TeV are, in fact, thought to be released now by RX~J1713.7-394 \cite{zira}. 
In the three regions a, b and d 
the high energy emission is clearly dominated by the radiation produced along the line of sight 
distance between 900 and 1100 parsecs, plotted 
in dashed lines in Figure \ref{fig66}. The gamma-ray emission from high latitude regions, such as 
region c, is instead dominated by the contribution from IC scattering of background electrons, 
almost at all energies. 
In regions closer to the Galactic Plane the emission from inverse Compton scattering of background electrons 
is subdominant at TeV energies, where runaway cosmic rays produce the enhanced emission. Therefore the regions 
where to look for the emission from runaway particles are low latitude regions of higher gas density. 
The $\gamma$-ray spectra show a peculiar concave shape, being soft at low energies and hard at high energies, which, 
as discussed in \cite{Gabici2009}, might be important for the studies of the spectral compatibility 
of GeV and TeV gamma ray sources.  The peculiar spectral 
and morphological features of the gamma-ray due to runaway CRs can be 
therefore revealed by combining the spectra and gamma-ray images provided by the Fermi and Agile telescopes 
at GeV and by present and future ground based detectors at TeV energies. 
As shown by the surveys of the Galaxy, published by Fermi 
at MeV-GeV energies, by HESS at TeV energies and at very high energies by the Milagro Collaboration, the 
various extended Galactic sources differ in spectra, flux and morphology. However, there is growing 
evidence for the correlation of GeV and TeV energy sources. These 
sources appear often spectrally and morphologically different at different energies, possibly 
due not only to the better angular resolution obtained by the instruments at TeV energies, 
but also to the energy dependence of physical processes, such as CR injection and CR 
diffusion. For this reason it is important to properly model what we expect to observe at different energies 
by conveying in a quantitave way all information by recognizing that the enviroment, 
the source age, the acceleration rate and history, all play a role in the physical 
process of injection and all have to be taken into account for the predictions. 
Figure \ref{fig2} shows the ratio of the hadronic gamma-ray emission due to total CR spectrum to that of the
background CRs for the entire region under consideration. In our modeling only CRs with energies above about 100 
TeV have left the acceleration site \cite{zira} and the morphology of the emission depends upon the 
energy at which one observes the hadronic gamma-ray emission. The
hadronic gamma-ray emission and the ambient gas distribution are correlated
if and only if the parent CRs have already been released by the SNR and
had time enough to diffuse into the ISM. The different 
spatial distribution of the emission is also due to the different energy-dependent 
diffusion coefficients, assumed in the three different panels, making it into a useful tool to investigate the highly 
unknown CR diffusion coefficient \cite{Casanova2010b,Gabici2010}. 
\begin{figure}
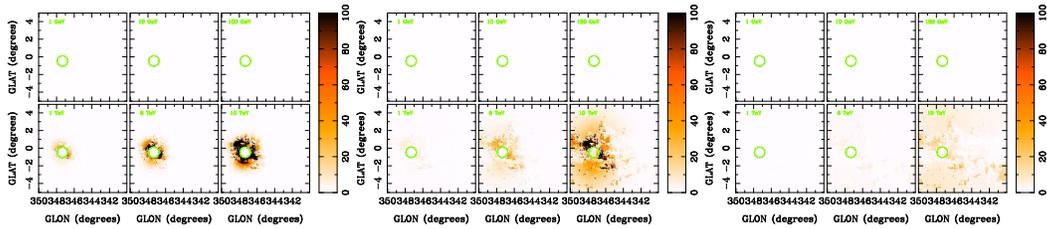

\centering
\includegraphics[height=0.3\textwidth,angle=270]{figur4a.eps}
\includegraphics[height=0.3\textwidth,angle=270]{figur4b.eps}
\includegraphics[height=0.3\textwidth,angle=270]{figur4c.eps}
\caption{Ratio of the emission due to the sum of background CRs and runaway CRs 
and background CRs only. The SNR is supposed to have exploded at 1 kpc distance from the Sun at ${347.3}^{\rm o}$
longitude and $-0.5^{\rm o}$ latitude 1600 years ago and to have started injecting the 
most energetic protons 100 years after the explosion \cite{zira}. 
The diffusion coefficient assumed within the region $340^{\rm
o}<l<350^{\rm o}$ and  $-5^{\rm o}<b<5^{\rm o}$ is 10$^{26}$ cm$^2$/s in
the left panel, 10$^{27}$ cm$^2$/s in the middle panel and 10$^{28}$ cm$^2$/s in
the right panel. In the three panels the ratio of the emission is shown for different energies 
from 1 GeV to 10 TeV.}\label{fig2}
\end{figure}

\section{Constraining the diffusion coefficient}\label{sec:w28}

\begin{figure}
\centering
  \includegraphics[width=0.9\textwidth]{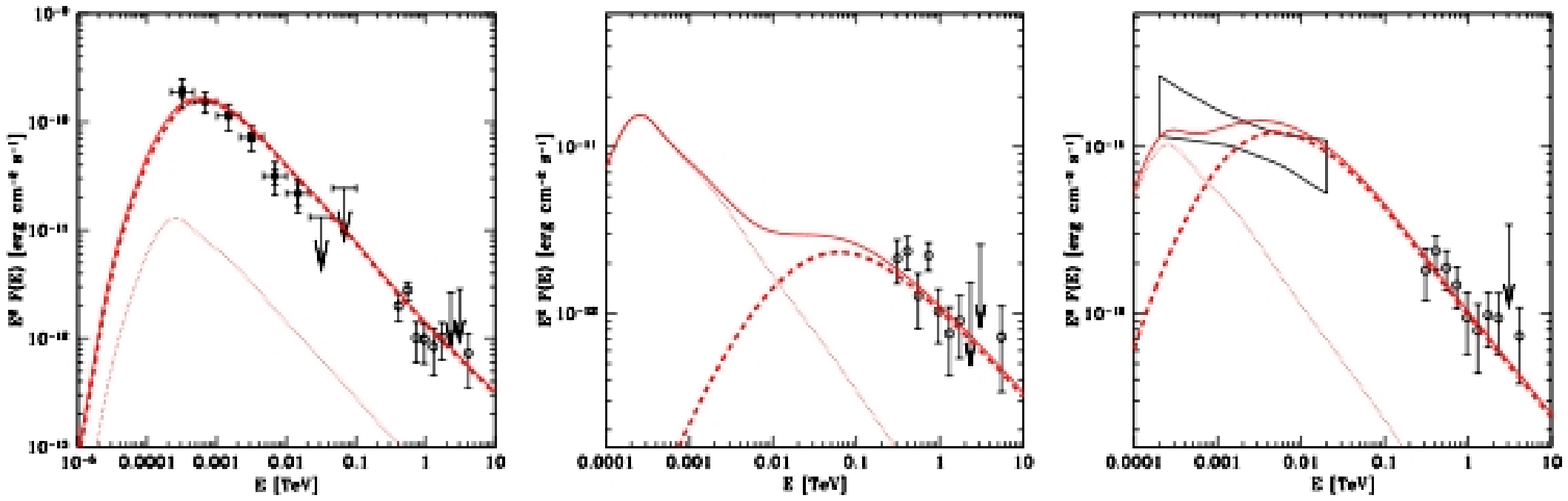}
\caption{Broad band gamma ray emission detected by FERMI
 and HESS (black data points) from the sources HESS J1801-233,
HESS J1800-240 A and B (left to right). Dashed lines represent the
 contribution to the predicted gamma ray emission from CRs that
   escaped W28, dotted lines show the contribution from the CR
    galactic background, and solid lines the total emission. The
        diffusion coefficient is assumed to be 6 $ \times {10}^{26}$ cm$^2$/s. \label{figureDiff}}
\end{figure}

Data on primary versus secondary CRs and on the residence time of long lived radioactive isotopes give us 
an idea of the average CR diffusion coefficient in the Galaxy, which is about ${10}^{28}$ cm$^2$/s. However, this does not 
exclude that the diffusion coefficient might vary locally (both in time and space). 
Figure \ref{figureDiff} shows Fermi and HESS spectra from the regions of the old SNR 
W28. The gamma ray observations can be explained by runaway CRs,
which have been accelerated in the past at the SNR shock, and subsequently escaped in the surrounding medium, only if the diffusion 
coefficient in the region surrounding the SNR is significantly supressed with 
respect to the average Galactic diffusion coefficient \cite{Gabici2010}.

\section{Molecular Clouds as Cosmic Ray Barometers}\label{sec:background}
\begin{figure*}
\begin{center}
\includegraphics[width=0.4\textwidth]{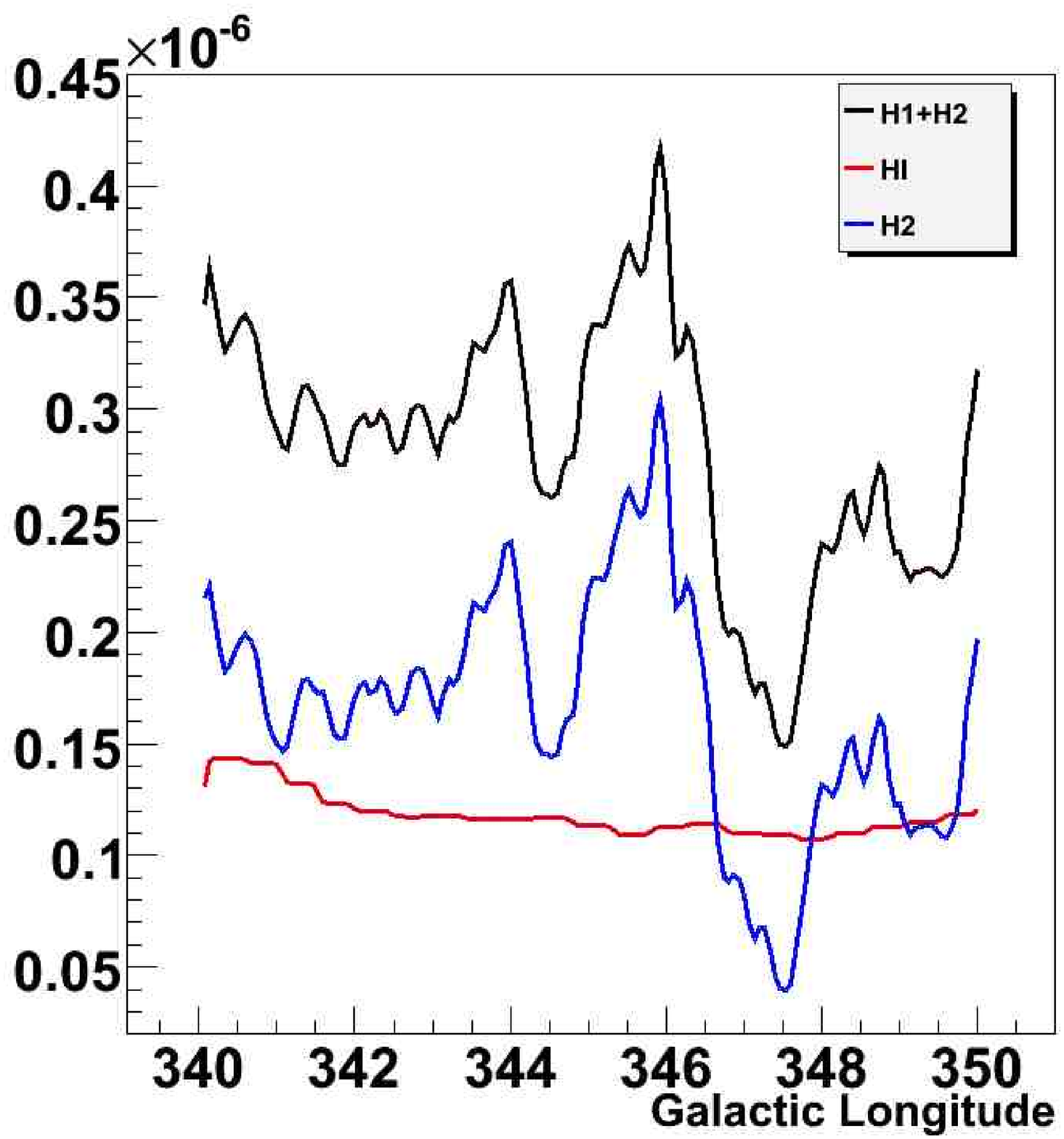}
\includegraphics[width=0.4\textwidth]{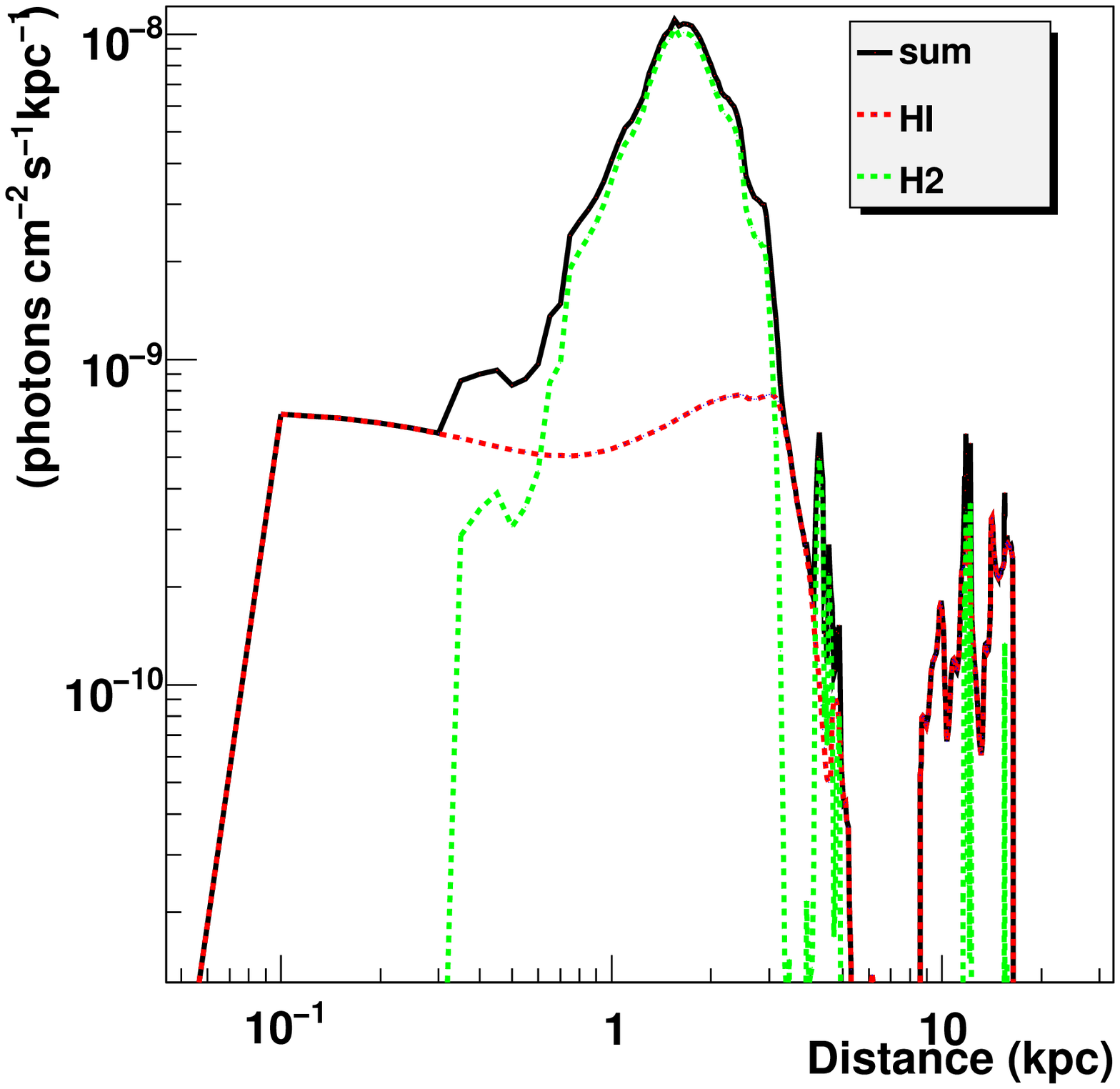}
\end{center}
\caption{ ({\it Left})The longitudinal profile of the $\gamma$-ray emission from the region $340^{\rm o}<l<350^{\rm o}$, 
integrated over the latitude range $-5^{\rm o}<b<5^{\rm o}$. 
 The dotted green lines are for the emission arising from CRs scattering off molecular hydrogen, the dashed red ones for the emission 
arising from 
CRs scattering off atomic hydrogen and the solid black ones for the sum. ({\it Right}) Flux above 1 GeV as function of the line of sight distance, 
produced by CRs interacting with the atomic and molecular gas. \cite{Casanova2010a} \label{fig4} }
\end{figure*}

We here describe a methodology which uses the emissivity of molecular clouds located far away from CR sources, 
so called {\it passive} molecular clouds, to probe the level
of the CR background \cite{Issa}. The longitudinal profile of the gamma-ray emission due to protons scattering off the atomic \cite{Kalberla} and molecular hydrogen \cite{Fukui1,Fukui2} 
from the region 
which spans Galactic longitude $340^\circ<l<350^\circ$ and Galactic latitude $-5^\circ<b<5^\circ$ is shown in Figure \ref{fig4} ({\it Left}). 
A peak in the emission at longitude of about $345.7^\circ$ close to the Galactic Plane is clearly visible, next to a 
dip in the longitude profile. While 
the atomic gas is generally broadly distributed along the Galactic Plane, the molecular hydrogen is less uniformly distributed and the peaks in the 
$\gamma$-ray longitude profiles correspond to the locations of highest molecular gas column density. The peaks in the 
longitudinal profile reveal the directions 
in the Galaxy where massive clouds associated with spiral arms are 
aligned along the line of sight. Fig. \ref{fig4} ({\it Right}) shows that the peak in the $\gamma$-ray emission from the direction ${345.7}^{\rm o}$ close to 
the Galactic plane is mostly produced within 0.5 kpc and 3 kpc distance from the Sun (in fact 85 percent of the emission 
is produced  in a region within 0.5 kpc and 3 kpc distance from the Sun). 
Thus the $\gamma$-ray emission from this direction provides a unique probe of the CR spectrum in 0.5-3 kpc \cite{Casanova2010a}. 
Given that the gamma-ray-emission from 
the molecular cloud depends only upon the total mass of the cloud, 
M, and its distance from the Earth, $d$, the CR flux, $\Phi_{CR}$, in the cloud is uniquely determined as
\begin{equation}
 \Phi_{CR} \propto \frac{{F_{\gamma}} \, d^2} {M} 
\label{eqn:fluxCloud}
\end{equation} 
where ${F_{\gamma}}$ is the integral gamma-ray flux from the cloud. 
Under the assumption that the CR flux in the cloud is equal to the locally observed CR flux, the calculated $\gamma$-ray flux from the cloud 
can be compared to the observed gamma-ray flux in order
to probe the CR spectrum in distant regions of the Galaxy. The detection of under-luminous clouds with 
the respect to predictions based on the CR flux at Earth would 
suggest that the local CR density is enhanced 
with respect to the Galactic average density. This would cast doubts 
on the assumption that the local CRs are produced only by distant sources, and 
that the CR flux and spectrum measured locally is representative
of the average Galactic CR flux and spectrum.

\section{Conclusions}\label{sec:conclusions}

The emission from the regions surrounding young SNR shells can 
provide crucial informations on the
history of the SNR acting as a CR source and important constraints 
on the highly unknown diffusion coefficient. Detailed modeling of the 
energy spectra and of the spatial distribution of the gamma-ray emission 
in the environment surrounding RX~J1713.7-3946 have been presented by 
using the data from atomic and molecular hydrogen surveys. Also, 
combining Fermi and HESS observations from the region around the old 
SNR 
W28 we have constrained the local diffusion 
coefficient, which we found to be significantly supressed with 
respect to the average Galactic diffusion coefficient. Furthermore we have presented a methodology 
to test the cosmic ray flux 
in discrete distant regions of the Galaxy by comparing the predicted and the 
measured gamma-ray flux from dense MC regions.

\end{document}